\documentclass[%
 reprint,
 prd,f
 amsmath,amssymb,
 aps,
 lengthcheck,
floatfix,
]{revtex4}

\usepackage{graphicx}
\usepackage{dcolumn}
\usepackage{bm}
\usepackage{hyperref}
\usepackage[utf8]{inputenc}
\usepackage{multirow}
\usepackage{soul}
\usepackage{float}
\usepackage{graphicx}
\usepackage{times}
\usepackage[normalem]{ulem}
\usepackage{color}
\usepackage{cellspace}
\usepackage[usenames,dvipsnames]{xcolor}

\usepackage{lipsum}

\newcommand{\be}{\begin{equation}}
\newcommand{\ee}{\end{equation}}
\newcommand{\bea}{\begin{eqnarray}}
\newcommand{\eea}{\end{eqnarray}}

\newcommand{\comment}[1]{}
\renewcommand\sout{\bgroup \color{red} \ULdepth=-.5ex \ULset}

\def\simge{\mathrel{\rlap{\raise 0.511ex
     \hbox{$>$}}{\lower 0.511ex \hbox{$\sim$}}}}
\def\simle{\mathrel{\rlap{\raise 0.511ex
      \hbox{$<$}}{\lower 0.511ex \hbox{$\sim$}}}}

\DeclareUnicodeCharacter{2212}{-}
\usepackage[none]{hyphenat}

\begin{document}

\title{Establishing  connection between  neutron star properties and nuclear matter parameters \\ through  a comprehensive multivariate analysis}

\author{\href{https://orcid.org/0000-0003-0103-5590}N. K. Patra$^{1}$\includegraphics[scale=0.06]{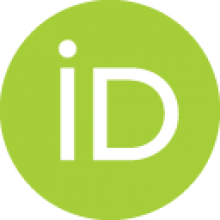}}
\email{nareshkumarpatra3@gmail.com}

\author{\href{https://orcid.org/0000-0002-4997-6544}Prafulla Saxsena$^{2}$\includegraphics[scale=0.06]{Orcid-ID.png}}
\email{prafulla1308@gmail.com}

\author{\href{https://orcid.org/0000-0001-5032-9435}B. K. Agrawal$^{3,4}$\includegraphics[scale=0.06]{Orcid-ID.png}
}
\email{bijay.agrawal@saha.ac.in}

\author{\href{https://orcid.org/0000-0002-9334-240X}T. K. Jha$^{1}$\includegraphics[scale=0.06]{Orcid-ID.png}}
\email{tkjha@goa.bits-pilani.ac.in}

\affiliation{$^1$Department of Physics, BITS-Pilani, K. K. Birla Goa Campus, Goa 403726, India}

\affiliation{$^2$ Malaviya National Institute of Technology, Jaipur, India}

\affiliation{$^3$Saha Institute of Nuclear Physics, 1/AF 
Bidhannagar, Kolkata 700064, India}  
\affiliation{$^4$Homi Bhabha National Institute, Anushakti Nagar, Mumbai 400094, India.}

\date{\today}

\begin{abstract}
 
We have attempted to mitigate the challenge of connecting the neutron star (NS) properties with the nuclear matter parameters that describe equations of state (EoSs). The efforts to correlate various neutron star properties with individual nuclear matter parameters have been inconclusive.
A Principal Component Analysis is employed as a tool to uncover the connection between multiple nuclear matter parameters and the tidal deformability as well as the radius of neutron stars within the mass range of $1.2-1.8M_\odot$. The essential EoSs for neutron star matter at low densities have been derived using both uncorrelated uniform distributions and minimally constrained joint posterior distributions of nuclear matter parameters. For higher densities ($\rho > 0.32$fm$^{-3}$), the EoSs have been established through a suitable parameterization of the speed of sound, which consistently maintains causality and gradually approaches the conformal limit. Our analysis reveals that in order to account for over 90\% of the variability in NS properties, it is crucial to consider two or more principal components, emphasizing the significance of employing multivariate analysis. To explain the variability in tidal deformability needs a greater number of principal components compared to those for the radius at a given NS mass. The contributions from iso-vector nuclear matter parameters to the tidal deformability and radius of NS decrease by $\sim$ 25\% with the increase in mass of NS from 1.2$M_\odot$ to 1.8$M_\odot$.
\end{abstract}


\maketitle
\section{Introduction}

The physics governing the dense matter within neutron stars (NSs) remains in mystery to this day \cite{Haensel2007}. These celestial objects contain matter upto a very high density at the center up to many times the nuclear saturation density ($\rho_0 \approx 0.16$fm$^{-3}$). The nuclear equation of state (EoS) plays a critical role in determining the structure of NSs. However,  a large number of speculative EoS models have been developed as a result of the lack of precise information about nuclear interactions at the densities found within the interior of neutron stars. The challenge lies in accessing matter at supra-nuclear densities, a realm beyond the reach of terrestrial experiments. Astrophysical observations provide essential knowledge for understanding the EoS of dense matter. The observation of neutron stars with mass $\sim$ 2$M_\odot$ \cite{Antoniadis:2013pzd, NANOGrav:2017wvv} established a lower limit on the maximum mass that an EoS must predict.

Significant progress has been achieved, especially since GW170817, in better understanding the nuclear matter equation of state by utilizing Bayesian statistical tools and various nuclear EoS models to analyze data from both astrophysical observations and terrestrial nuclear experiments \cite{ Malik2018, Chatziioannou:2020pqz, Baiotti:2019sew, Li:2021thg, Lim:2019som, Raithel:2019uzi, Fattoyev2018a, Forbes:2019xaz, Landry2019, Piekarewicz2019, Biswas2020, Abbott2020, Thi2021, Patra:2022lds, Lovato:2022vgq, Sorensen:2023zkk}. The behavior of dense matter in the neutron stars has a significant impact on the characteristic  GW signals that come from the merging of binary neutron stars \cite{Abbot2018, Abbott2019, GW170817, GW190814}. Consequently, discovering these gravitational waves offers a rare chance to limit the nuclear matter parameters (NMPs) that define the equation of state. The tidal deformability of the NS is an intriguing property inferred by GWs, and this trait has been investigated for a variety of models for EoSs\cite{Flanagan:2007ix, Hinderer:2007mb, Damour:2009vw, Patra:2022yqc}. The precise X-ray observations of hot spots on pulsars by Neutron star Interior Composition Explorer (NICER) \cite{Miller:2019cac, Riley:2019yda, Miller:2021qha, Riley:2021pdl} and observations of gravitational waves from NS mergers by LIGO/VIRGO \cite{LIGOScientific:2014pky, VIRGO:2014yos} are expected to put constraints on the tidal deformability and radius.

Numerous initiatives have been made in the field of neutron star study to investigate the fascinating relationships between the star's tidal deformability, radius, and the crucial nuclear matter parameters driving the density-dependent symmetry energy~ \cite{Alam:2016cli, Carson:2018xri, Malik2018, Tsang:2019vxn, Guven:2020dok, Malik:2020vwo, Tsang:2020lmb, Malik_book, Reed:2021nqk, Pradhan:2022vdf, Pradhan:2022txg, Beznogov:2022rri}.  Recent studies explore the impact of constraining the low-density equation of state using more than 400 mean-field models arising from the non-relativistic Skyrme interactions and relativistic Lagrangians governing nucleon interactions through $\sigma$, $\omega$, and $\rho$ mesons \cite{Carlson:2022nfb}. A strong correlation between the tidal deformability of canonical $1.4 M_\odot$ mass neutron stars and the slope of the symmetry energy at the saturation density ($\rho_0$) has been revealed by this group of models, which are adept at capturing the characteristics of symmetric and asymmetric finite nuclei.
Although earlier approaches frequently relied on extracting these nuclear matter parameters from uncorrelated uniform or Gaussian distributions, their relationships to the characteristics of neutron stars were generally weak/moderate.
Thus, determining the key nuclear matter parameters that have the greatest impact on the neutron star's properties, such as tidal deformability and radius at canonical mass 1.4 $M_\odot$ is still inconclusive \cite{Carson:2018xri, Pradhan:2022vdf, Pradhan:2022txg, Malik2018, Tsang:2019vxn, Malik:2020vwo, Tsang:2020lmb, Malik_book, Reed:2021nqk, Kunjipurayil:2022zah}.

The current observational lower limit on the maximum mass of neutron stars, around $2M_\odot$, suggests that the central density of a neutron star with a canonical mass of 1.4$M_\odot$ could be within the range $\sim 2–3\rho_0$ \cite{Hebeler:2013nza, Li:2005sr, Patra:2022yqc}. Understanding the behavior of equations of state around this nuclear saturation density is crucial for determining the characteristics of these neutron stars. The existence of a strong connection between the radius of neutron stars in the mass range of 1-1.4$M_\odot$ and the pressure of $\beta-$equilibrated matter at densities  $1-2\rho_0$ has been shown \cite{Lattimer:2000nx}. Similarly, studies have extended to the tidal deformability, revealing a strong correlation with pressure at $2\rho_0$ \cite{Tsang:2019vxn, Tsang:2020lmb, Patra:2022yqc, Imam:2023ngm}. Despite variations in neutron star properties, their dependence on the pressure of $\beta$-equilibrated matter at twice the saturation density remains consistent. This robustness might stem from the dependence of pressure on several nuclear matter parameters that describe symmetric nuclear matter and the density-dependent symmetry energy. Recent works have established empirical relationships between the radius and tidal deformability with multiple nuclear matter parameters \cite{Imam:2023ngm, Richter:2023zec}. Therefore, a multivariate analysis could prove essential in understanding the contributions of different nuclear matter parameters across a wide range of neutron star masses.

In the present work, we perform a Principal Component Analysis (PCA) to investigate the connection between multiple nuclear matter parameters and the tidal deformability as well as the radius of neutron stars across a wide range of masses. The analysis is carried out using two different sets of  EoSs. The EoSs at low densities have been derived using the uncorrelated uniform and joint posterior distributions of nuclear matter parameters. The joint posterior distributions of nuclear matter parameters are obtained by imposing the constraints on the low-order nuclear matter parameters determined by the experimental data on the bulk properties of finite nuclei together with the pure neutron matter (PNM) EoS from a precise next-to-next-to-next-to-leading-order (N$^{3}$LO) calculation in chiral effective field theory \cite{Hebeler:2013nza, Lattimer:2021emm} within the Bayesian Inference. The EoSs at high density ($\rho>2\rho_0$) are constructed by imposing the causality condition on the speed of sound and are independent of compositions of NS matter. Our analysis reveals that more than one principal component is necessary to appropriately describe the NS properties, such as tidal deformability and radius. The role of iso-scalar nuclear matter parameters becomes increasingly important with neutron star mass.

The paper is organized as follows.  We briefly outline our methodology in Sec.\ref{methe}. The results for multivariate analysis of the neutron key properties are discussed in detail in Sec. \ref{results}. The summary and outlook are presented in Sec. \ref{summary}. 

\section{Methodology} 
\label{methe}
We discuss the construction of equations of state at low and high densities, using Taylor expansion for low-density EoS below 2$\rho_0$, and the high-density EoS is constructed by adjusting the speed of sound to maintain causality.

\subsection{Equation of State at Low and High-densities}
\label{EoS-construct}
The energy per nucleon at a given density $\rho$ and asymmetry $\delta$ is expressed as follows, using a parabolic approximation:
\bea
E(\rho, \delta) &=&  E(\rho,0)+E_{\rm sym}(\rho)\delta^2
+..., \label{eq:EoS}
\eea
where $\delta = \frac{\rho_n -\rho_p}{\rho}$ is calculated using $\beta$-equilibrium and the charge neutrality criteria. Using individual nuclear matter parameters, symmetric nuclear matter energy  $E(\rho,0)$ and density-dependent symmetry energy $E_{\rm sym}(\rho)$ are expanded around $\rho_0$ as 
\cite{Chen:2005ti, Chen:2009wv, Newton:2014iha, Margueron:2017eqc, Margueron:2018eob},
\bea
E(\rho, 0)&=&	e_0+\frac{1}{2}K_0\left (\frac{\rho-\rho_0}{3\rho_0}\right )^2
+\frac{1}{6} Q_0 \left(\frac{\rho-\rho_0}{3\rho_0}\right )^3, \label{eq:SNM_T} \\ 
E_{\rm sym}(\rho) &=&	J_0 + L_0\left (\frac{\rho-\rho_0}{3\rho_0}\right )
+\frac{1}{2}K_{\rm sym,0}\left (\frac{\rho-\rho_0}{3\rho_0}\right )^2 \nonumber \\  
&+& \frac{1}{6}Q_{\rm sym,0}\left (\frac{\rho-\rho_0}{3\rho_0}\right )^3. \label{eq:sym_T}  
\eea 
In the above Eqs. (\ref{eq:SNM_T},\ref{eq:sym_T}), $e_0 $ is the binding energy per nucleon, $K_0$ is the incompressibility coefficient, $J_0$ is the symmetry energy coefficient, its slope parameter $L_0$, $K_{\rm sym,0}$ is the symmetry energy curvature parameter, $Q_0$ [$Q_{\rm sym,0}$] are the skewness parameter of  $E (\rho,0) $  $[E_{\rm sym}(\rho)]$.

In order to build the EoS beyond the density $2\rho_0$, we use the causality condition on the speed of sound. The low-density component of the EoS ($\rho <2\rho_0$) links to the high-density part, ensuring that the sound velocity asymptotically approaches the conformal limit ($c_s^2$ = $\frac{1}{3}c^2$) and never surpasses the speed of light. The sound velocity for $\rho>2\rho_0$ is given as\cite{Tews:2018kmu},
  
\bea
\frac{c_s^2}{c^2} &=& \frac{1}{3} - c_1 exp\left[{-\frac{(\rho-c_2)^2}{n_b^2}}\right] + h_p exp\left[-\frac{(\rho-n_p)^2}{w_p^2}\right]\nonumber\\
&& \left[1 + erf(s_p \frac{\rho-n_p}{w_p})\right]. \label{eq-vs}
\eea 
where the peak height $h_p$ defines the maximum speed of sound, the position $n_p$ determines the density in the area where it occurs, the width of the curve is controlled by the $w_p$ and $n_b$ variables, and the shape or skewness parameter $s_p$. The continuity of the speed of sound and its derivative at the transition density $\rho_{\rm tr}$ determines the parameters $c_1$ and $c_2$ for a given value of $n_b$. The values of $n_b$, $h_p$, $w_p$, and $n_p$ are taken from a uniform distribution with ranges of $0.01-3.0$ fm$^{-3}$, $0.0-0.9$, $0.1-5.0$ fm$^{-3}$, and $(\rho_{\rm tr} + 0.08) - 5.0$ fm$^{-3}$, respectively \cite{Tews:2018kmu}. Since $s_p$ barely affects the stiffness of EoS, we set it equal to zero for all of our calculations.

We start with the transition density ($\rho_{\rm tr}$), where the energy density ($\epsilon(\rho_{\rm tr})$), the pressure (P($\rho_{\rm tr}$)), and the derivative of the energy density ($\epsilon^\prime(\rho_{\rm tr})$) are known. Assuming a step size of ($\Delta \rho=0.001$ fm$^{-3}$), the following formula is used to produce consecutive values of $\epsilon$ and P:

\bea
 \rho_{i+1} &=& \rho_i + \Delta \rho \label{eq-rhoE},\\
 \epsilon_{i+1} &=& \epsilon_i + \Delta \epsilon \nonumber\\
                &=& \epsilon_i + \Delta \rho \frac{\epsilon_i + P_i}{\rho_i}\label{eq-engE},\\
P_{i+1} &=&P_i + c_s^2(\rho_i) \Delta \epsilon  \label{eq-preE}.              
\eea
where i = 0 is the transition density $\rho_{\rm tr}$. In Eq. (\ref{eq-engE}), $\Delta \epsilon$ was assessed using the thermodynamic relation $ P = \rho \partial \epsilon/{\partial \rho}-\epsilon$ valid at zero temperature. After generating the EoS, the Tolman-Oppenheimer-Volkoff (TOV) equations are solved to determine NS properties, including tidal deformability and radius as a function of mass.

\subsection{Principal Component Analysis}\label{PCA}
Principal Component Analysis (PCA) is an important statistical technique for analyzing multivariate data \cite{Wold1987, Aflalo2017, Al-Sayed:2015voa, shlens2014, Liu:2020ely, Acharya:2021lbe, Ali2023}. The PCA method analyzes the data of multiple dependent correlated variables to capture shared variation.
The objectives of PCA \cite{Abdi2010} are (i) Extract the most crucial information from the data, (ii) Reduce the dimensionality of the data by only maintaining the essential information, (iii) Examine the composition of the observations and variables, as encountered in literature.
In the present work, we considered  $K_0$, $Q_0$, $J_0$, $L_0$, $K_{\rm sym0}$ and  $Q_{\rm sym0}$ as our variables often referred to as features and the neutron stars properties such as tidal deformability and radius as our target variables. PCA  can verify our earlier hypothesis of selecting considered variables and prove the significance of these variables. 
In the process of PCA, the method computes Principal Components (PCs) that can be considered as new variables in other dimensions. PCs are the composition of linear combinations of original variables to capture the shared variation patterns. These PCs account for the amount of variation captured in data and return relative scores as eigenvalues in a sorted manner. So, the largest eigenvalue associated with the first PC captures the largest variance, and so on. 

The methodology of PCA analysis is composed of the given steps:
(1) The covariance matrix is calculated from the given data. The covariance matrix measures the relationships between pairs of variables.
(2) Eigenvalue decomposition on the covariance matrix is carried out to obtain the eigenvectors and eigenvalues. 
(3) The eigenvectors represent the principal components, and the corresponding eigenvalues indicate the principal components' proportional variance captured.

To calculate the covariance matrix, we prepare our data $\bf{X}$ as a $I \times J$ matrix. We have $'I'$ samples that are represented by $'J'$ variables. We must standardize the data set by removing the mean and dividing by the standard deviation of each $\bf{X}$ column. This yields a standardized data matrix, such as $\bf{\dot{X}}$. The elements of the covariance matrix are determined as,
\bea
{{\bf{C}_{ij}}} &=& \frac{1}{n} \sum^n_{k=1} {\dot{\rm \bf{X}}_{ik}}{\dot{\rm \bf{X}}_{jk}}, 
\eea
where ${i,j}$ denotes the variables/features, and $\bf{\mathit{n}}$ runs over all the samples. 
The correlation matrix is used to determine the eigenvalues and eigenvectors. The order of eigenvalues provides the importance of the eigenvector. The most important principal component (PC1) is the eigenvector with the greatest corresponding eigenvalue. PC1 captures the highest variance among the data. The second component (PC2) must be orthogonal to the first component and capture the second-highest variance. In PC space, factor scores indicate observations (samples), which is the projection of data along with the PC components. The factor score matrix $\bf{F}$ is defined as,
\bea
\bf{F}= \bf{\dot{X}}\bf{V}. \label{eq-factor}
\eea
The matrix $\bf{V=\dot{X}A}$ is called a factor loading matrix, and matrix $\bf{A}$  contains eigenvectors. Matrix $\bf{F}$ gives the projections of observations on primary components, making it a projection matrix. 
Each PC's contribution to each original variable shows the captured variance. The importance of PC is decided based on the corresponding order of eigenvalue; hence, the contribution of the original variable also depends on the order of PC priority. 

\section{Multivariate Analysis of the NS Properties} \label{results}

A multivariate analysis is performed to investigate the connection of the NS properties with
multiple nuclear matter parameters. The required EoSs for a given
set of nuclear matter parameters are obtained as outlined in the previous section. These EoSs satisfy the thermodynamic stability and causality condition and yield the maximum mass larger than $2M_\odot$. Two different sets of EoSs are obtained, which correspond to different distributions of nuclear matter parameters as described
below. The tidal deformability and radius over a wide range of NS masses are calculated from the solutions of TOV equations. These exact values of NS properties are then fitted to the linear functions of nuclear matter parameters and employed for the PCA. The PCA allows us to appropriately identify the important nuclear matter parameters required to explain the variations of a given  NS property.

\subsection{Distributions of Nuclear Matter Parameters}
\begin{table}[htp]
\centering
\caption{\label{tab1} The lower (Min.) and upper (Max.) bounds of the uniform distributions (D1) along with the means ($\mu$) and uncertainties ($2\sigma$) of marginalized distributions (D2) for each nuclear matter parameter except for $e_0$, which is kept fixed to -16.0 MeV are listed.}
\begin{ruledtabular} 
\begin{tabular}{ccccc}
\multirow{2}{*}{NMPs} & \multicolumn{2}{c}{D1}  &\multicolumn{2}{c}{D2} \\ [1.5ex] 
  \cline{2-5}
{(in MeV)}    & Min. & Max. & $\mu$ & 2$\sigma$ \\[1.5ex] 
\cline{1-5}    
 $K_0$ & 200 & 290 & 242 & 45 \\[1.5ex]
 $Q_0$ & -500 & 450 & -25 & 466 \\[1.5ex]
 $J_0$ & 30 & 35 & 32.2 & 2 \\[1.5ex]
 $L_0$ & 30 & 80 & 54.2 & 24 \\[1.5ex]
 $K_{\rm sym,0}$ &-300 & 100 & -89 & 180 \\[1.5ex]
 $Q_{\rm sym,0}$ & 0 & 1000 & 772 & 700 \\[1.5ex] 
\end{tabular}
\end{ruledtabular}
\end{table}

\begin{figure*}[htp]
\centering
\includegraphics[width=\textwidth]{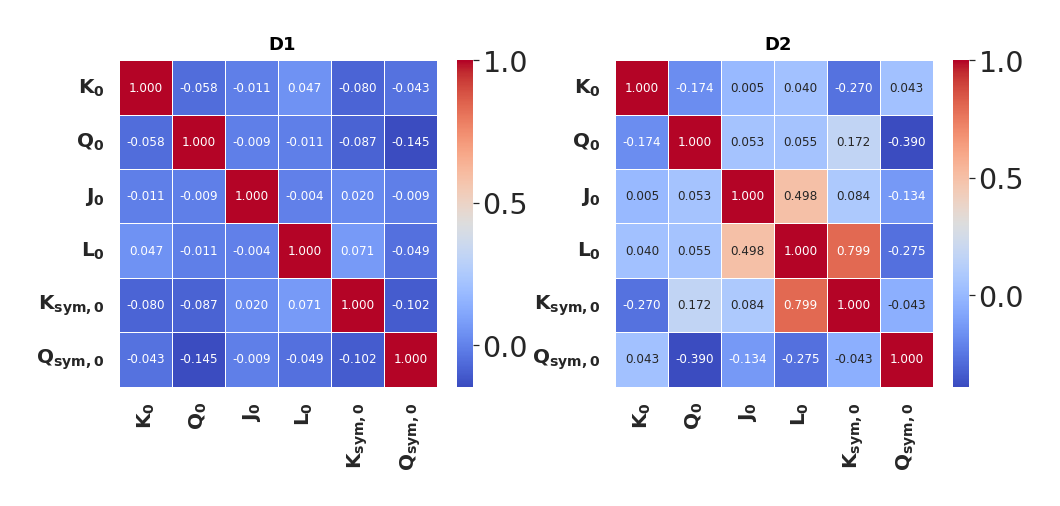}
\caption{(Color online) The correlation among various nuclear matter parameters is shown. The results for the left and right panels are for uncorrelated (D1) and correlated (D2) distributions of nuclear matter parameters, respectively. Color codes indicate the values of Pearson’s correlation coefficients among the various NMPs.}\label{fig1}
\end{figure*} 

We have generated two distinct sets of EoSs that correspond to different distributions of nuclear matter parameters.  {One of these sets is based on uncorrelated uniform distributions of nuclear matter parameters, and the other one is obtained from their joint posterior distribution. The joint posterior distribution of the nuclear matter parameters is taken from Ref.\cite{Patra:2023jvv}, which was obtained by imposing minimal constraints that include some selected basic nuclear matter properties at the saturation density and the EoS for the pure neutron matter at low densities from  N$^3$LO calculation in the chiral effective field theory \cite{Hebeler:2013nza, Lattimer:2021emm}. The conditions of thermodynamic stability, causality speed of sound on the EoS, and the resulting maximum mass of neutron star larger than 2$M_\odot$ were also imposed. Some of the nuclear matter parameters are correlated due to minimal constraints from the chiral effective field theory applied within the Bayesian statistical method \cite{Gelman2013, Buchner2014, Ashton2019}. The marginalized distributions of each nuclear matter parameter are derived from the joint posterior distribution in order to put the bounds on their uniform distributions. The bounds on uniform distributions of nuclear matter parameters roughly the 90\% confidence intervals of the corresponding marginalized distributions.}  In Table \ref{tab1}, we have listed the lower and upper bounds of the uniform distributions, as well as the means and variances of the marginalized distributions for each nuclear matter parameter. The iso-scalar nuclear matter parameters, the binding energy per nucleon $e_0$, and the saturation density $\rho_0$ for symmetric nuclear matter remain fixed at -16.0 MeV and 0.16 fm$^{-3}$, respectively. References \cite{Patra:2023jvv, Kunjipurayil:2022zah} have shown that the dependence of neutron star properties on individual nuclear matter parameters are sensitive to the choice of the distributions of nuclear matter parameters. We would like to explore how multivariate analysis is sensitive to different distributions of these nuclear matter parameters.

We have generated  10,000 samples of nuclear matter parameters for each of the uniform uncorrelated and joint posterior distributions, hereafter referred to as  D1 and D2, respectively.
{ For each set of nuclear matter parameters, the equations of the state beyond $2\rho_0$ are obtained for a random set of the speed of sound parameters $n_b, h_p, w_p,$ and $n_p$ as discussed in previous Sec. \ref{EoS-construct}.}
Out of these, around 2,500 samples from both distributions have been selected after applying the filters mentioned earlier.
In Fig. \ref{fig1}, we have illustrated the correlations among nuclear matter parameters for both D1 and D2. The distribution D1 displays poor correlations among the nuclear matter parameters, whereas D2 reveals stronger correlations among some of the parameters due to the minimal constraints, such as the notable correlation coefficient r($L_0$, $K_{\rm sym,0}$) = 0.80. The significant correlations between $L_0$ and $K_{\rm sym,0}$ have been previously documented as well \cite{Patra:2022yqc, Patra:2023jvv}.

\subsection{Fit of NS Properties to Nuclear Matter Parameters}

\begin{figure}[htp]
\centering
\includegraphics[width=0.5\textwidth]{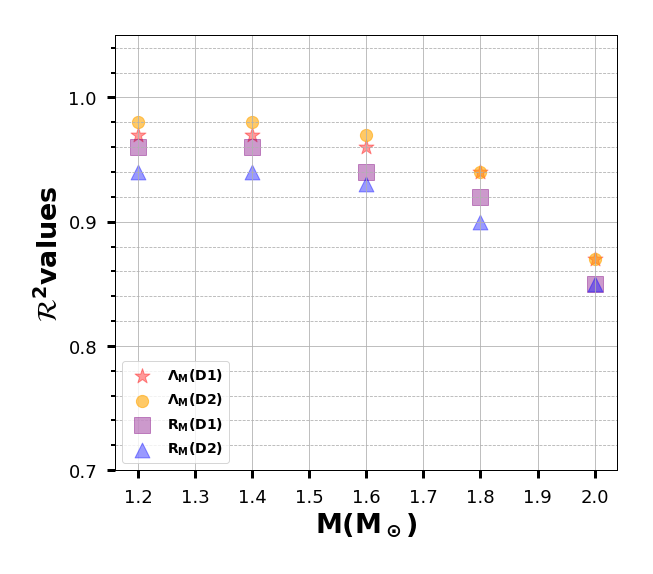}
\caption{(Color online)
The $\mathcal{R}^2$ values of the neutron star properties, such as tidal deformability and radius in the mass range of $1.2-2 M_\odot$, are shown. The different color symbols correspond to different NS properties and the data.}\label{fig2}
\end{figure}

The EoSs for $\beta$-equilibrated charge neutral matter in the density range $0.5\rho_0$ to $2\rho_0$ are constructed using Taylor expansion with the D1 and D2 distributions of nuclear matter parameters.  Each of these low-density EoS is smoothly joined by the EoSs
that satisfy causality conditions as given by Eqs. (\ref{eq-vs}-\ref{eq-preE}). The EoS for outer and inner crusts for density ranges $\rho<0.5\rho_0$ is used as follows. The inner crust EoS is polytropic \cite{Carriere:2002bx}, 
\bea p(\varepsilon)= \alpha + \beta \varepsilon^{\frac{4}{3}}. \label{eq-ic} 
\eea
The inner crust EoS is matched with the outer crust at one end and the outer core at the end by appropriately adjusting the values of the coefficients $\alpha$ and $\beta$. 
The inner crust's EoS affects the Love number $k_{2}$ and compactness parameter, but not the tidal deformability parameter \cite{Piekarewicz2019}. After determining the core and crust EoSs,  the neutron star mass, radius, and tidal deformability for a given central pressure can be computed using TOV equations.  In order to demonstrate our approach,  we use a linear fit function  of nuclear matter parameters as listed in Table~\ref{tab1} to calculate the tidal deformability ($\Lambda_M$) and radius ($R_M$) for a given NS mass as,
\bea
{\Lambda_M} &=& \sum_{i} W_i P_i + b=\sum_{i}\Lambda_i + b \label{eq-lam}\\
{R_M} &=& \sum_{i} W^\prime_i P^\prime_i + b^\prime=\sum_{i}R_i + b^\prime\label{eq-rad}
\eea
where $W_i$ and $W^\prime_i$ are the weight factors of given nuclear matter parameters. The $b$ and $b^\prime$ are the biases. The $P \in$ \{ $K_0$, $Q_0$, $J_0$, $L_0$, $K_{\rm sym,0}$, $Q_{\rm sym,0}$\} and M stands for the neutron star masses.  The $\Lambda_i$ and $R_i$ in the right hand side of above Eqs.(\ref{eq-lam}) and (\ref{eq-rad}) correspond to the $W_i  P_i$ and $W^\prime_i  P^\prime_i$, respectively. We assess the quality of fit for our regression model using the $\mathcal{R}^2$ value as a measure  \cite{Dennis2002}. The $\mathcal{R}^2$ value ranges from 0 to 1, where 0 signifies no variability in the dependent variable, and 1 represents complete variance. The  $\mathcal{R}^2$ close to unity indicates that the  NS properties from the solutions of TOV equations are almost equal to the ones obtained by their linear fit to the nuclear matter parameters. In Fig. \ref{fig2}, we present the $\mathcal{R}^2$  values obtained from fitting the tidal deformability and radius of neutron stars within the mass range of $M=1.2 - 2.0 M_\odot$. As the neutron star mass increases, the associated $\mathcal{R}^2$ value tends to decrease. Specifically, the $\mathcal{R}^2$ values for the fitted properties of neutron stars with a mass of 2$M_\odot$ fall below 0.9. Consequently, in what follows,  we focus on the multivariate analysis of neutron stars within the mass range of $1.2 - 1.8 M_\odot$, guided by the $\mathcal{R}^2$ value considerations. 

\begin{figure*}[htp]
\centering
\includegraphics[width=0.9\textwidth,height=10cm]{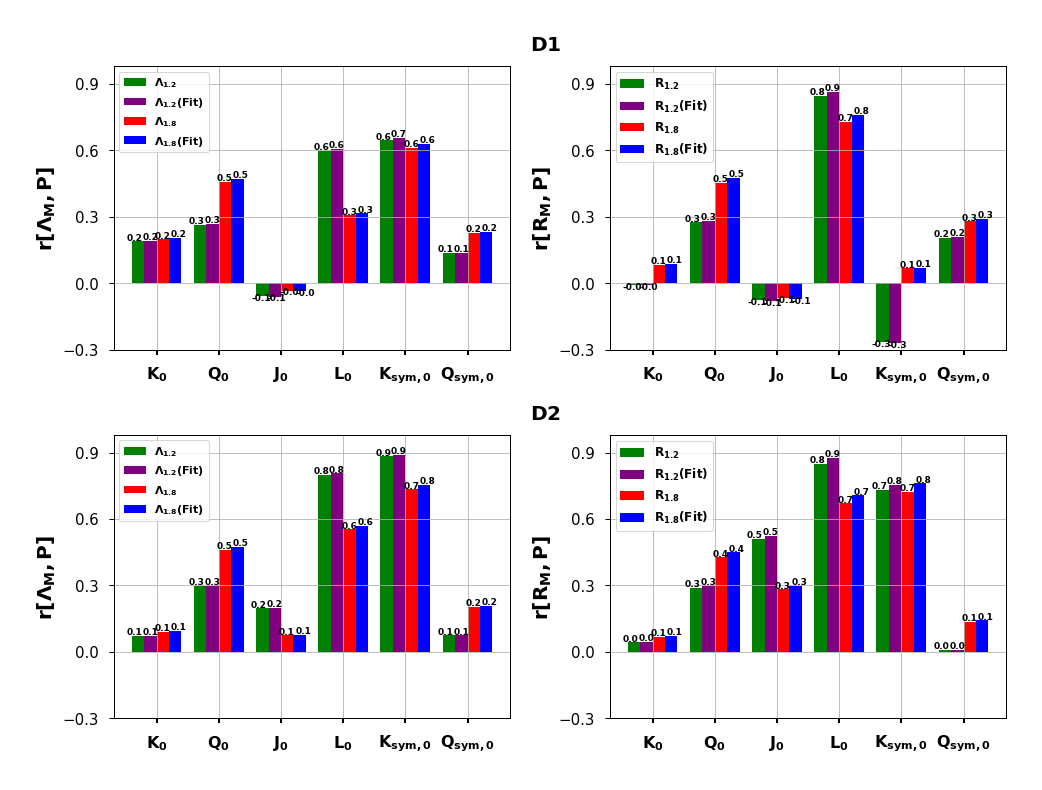}
\caption{(Color online) The bar plot shows correlation coefficient values of NS properties with nuclear matter parameters only for mass 1.2 and 1.8 $M_\odot$. For the comparison, the values of correlation coefficients from the fitted model are shown with different color bars. The upper and lower panels correspond to uncorrelated (D1) and correlated (D2) NMPs distributions.}\label{fig3}
\end{figure*}

In Fig.\ref{fig3}, we display the correlation coefficient values obtained for exact values of tidal deformability and radius with individual nuclear matter parameters for the masses 1.2 and 1.8  $M_\odot$  by green and red bars, respectively. We also juxtapose similar results obtained from the fitted values of tidal deformability and radius through Eqs. (\ref{eq-lam}) and (\ref{eq-rad}) for the masses 1.2 and 1.8  $M_\odot$  by purple and blue  bars, respectively. The results are presented for two different distributions of NMPs as indicated by labels  D1 (Upper) and D2 (Lower).  This visual representation demonstrates that all the correlation coefficient values remain consistent even after the fitting process. The application of the fit does not alter these correlation trends. The correlations of the tidal deformability with NMPs are qualitatively similar for the D1 and D2 cases. In general, the D2 distributions of NMPs result in stronger correlations. The correlations of the radius with the NMPs appear to be sensitive to the choice of their distributions. In particular, the radius is very weakly correlated with $K_{\rm sym,0}$ for the D1 case and shows reasonably strong correlations for the D2 case. It is also interesting to note that the correlations of tidal deformability and radius with  $Q_0$ increase somewhat with an increase in NS mass.

\subsection{Principal Component Analysis of NS Properties}
The method of PCA is implemented by following the Sec.\ref{PCA} and Appendix \ref{method-pca}. It is comprised of the following steps: (1) Construction of Covariance matrix 
(2) Diagonalization of the covariance matrix.
(3) The principal components' proportionate variance captured by the eigenvectors and eigenvalues. 
First, we calculate the values of weights $W_i$s and $W^\prime_i$s appear in Eqs. (\ref{eq-lam}) and (\ref{eq-rad}). The PCA is performed using $\Lambda_i$s and $R_i$s as features corresponding to the target variables $\Lambda$ and R, respectively. Then, $6\times 6$ covariance matrices for the tidal deformability and radius of NS are constructed for a given mass. The eigenvalues and the corresponding eigenvectors obtained by diagonalizing the covariance matrices are arranged in descending order. The most important principal component, PC1 corresponds to the eigenvector with the highest eigenvalue.  The succeding eigenvalues and the eigenvectors are labeled as  PC2, PC3, etc.  The nuclear matter parameter having the largest contribution to the eigenvector for PC1 is the most dominant one in determining the  NS property in consideration. Likewise, the eigenvectors associated with remaining PCs together with eigenvalues, can be used to identify other important nuclear matter parameters for the NS properties.

\begin{table*}[htp]
\caption{\label{tab2} The normalized eigenvalues associated with the principal components are listed for the NS properties with masses range $1.2-1.8M_\odot$. D1 and D2 correspond to the uncorrelated and correlated distributions of the nuclear matter parameters.
}
\centering
\setlength{\tabcolsep}{6.5pt}
\renewcommand{\arraystretch}{1.4}
\begin{ruledtabular}  
\begin{tabular}{cccccccc|cccccc}
\multirow{2}{*}{NS} & \multirow{2}{*}{$\frac{M}{M_\odot}$} &  \multicolumn{6}{c}{D1} & \multicolumn{6}{c}{D2} \\
 \cline{3-14}
&  & PC1 & PC2 & PC3 & PC4 & PC5 & PC6 & PC1 & PC2 & PC3 & PC4 & PC5 & PC6 \\[1.5ex]
\cline{3-14}
\hline
\multirow{5}{*}{{$\Lambda_M$}}& 1.2 & 1.00 & 0.62 & 0.32 & 0.17 & 0.13 & 0.01 & 1.00 &0.35 & 0.23 & 0.13 & 0.04 &0.00  \\[1.5ex]
&1.4 & 1.00 & 0.40 & 0.35 & 0.20 & 0.14 & 0.00 & 1.00 & 0.52 & 0.27 & 0.16 & 0.04 & 0.00 \\[1.5ex]
&1.6 & 1.00 & 0.53 & 0.25 & 0.22 & 0.15 & 0.00 & 1.00 & 0.67 & 0.33 & 0.20 & 0.03 & 0.00  \\[1.5ex]
&1.8 & 1.00 & 0.70 & 0.31 & 0.16 & 0.13 & 0.00 & 1.00 & 0.71 & 0.36 & 0.21 & 0.01 & 0.00  \\[1.5ex]
\multirow{5}{*}{{R$_M$}}& 1.2 & 1.00 & 0.12 & 0.10 & 0.07 & 0.01 &0.00 & 1.00 &0.18 &0.09 & 0.02 & 0.00 &0.00\\[1.5ex]
&1.4 & 1.00 & 0.19 & 0.12 & 0.03 & 0.00 & 0.00 & 1.00 & 0.29 & 0.13 & 0.01 & 0.00 & 0.00\\[1.5ex]
&1.6 & 1.00 & 0.31 & 0.18 & 0.01 &0.00 &0.00 & 1.00 & 0.46 & 0.20 &0.01 & 0.00 &0.00\\[1.5ex]
&1.8 & 1.00 & 0.54 & 0.29 & 0.03 & 0.02 &0.00 &1.00& 0.70& 0.32& 0.04& 0.00&0.00\\[1.5ex]

\end{tabular}
\end{ruledtabular}
\end{table*}

\begin{figure*}[htp]
\centering
\includegraphics[width=0.9\textwidth,height=12cm]{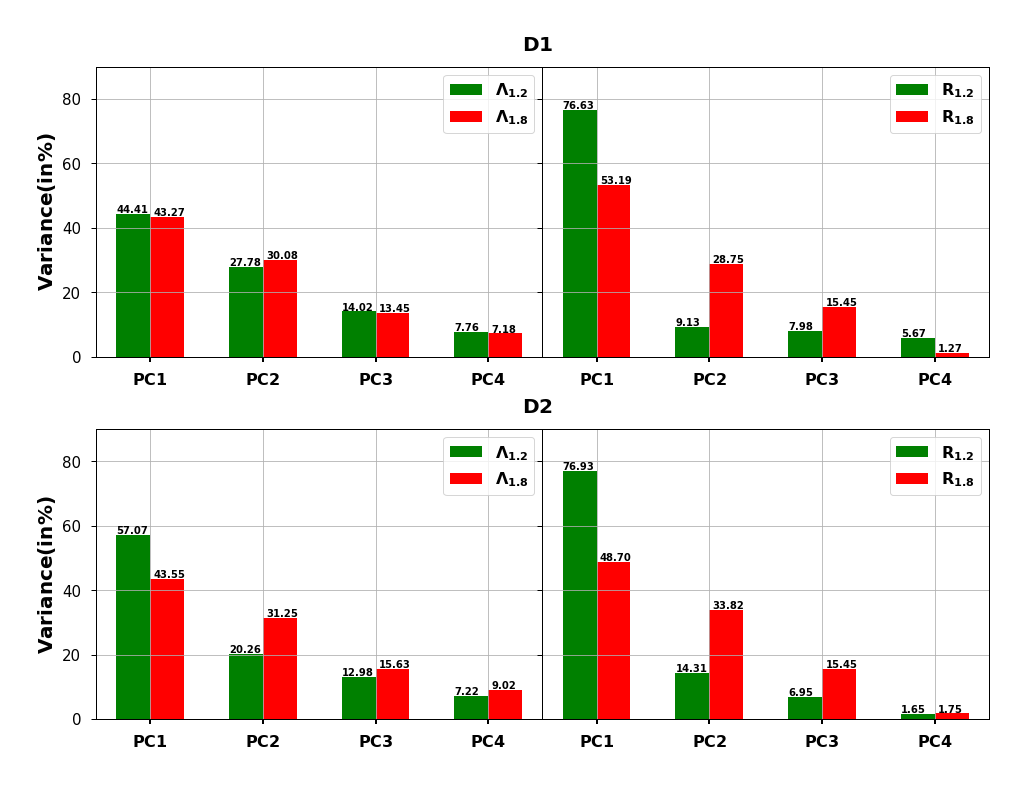}
\caption{(Color online) The variance (in \%) of different principal components corresponding to the NS properties is displayed. The results in the upper and lower panels represent the uncorrelated (D1) and correlated (D2) nuclear matter parameter distributions, respectively. }\label{fig4}
\end{figure*}
In Table \ref{tab2}, we provide a list of eigenvalues that were obtained by diagonalizing the covariance matrix for a specific property and mass of a neutron star. The highest eigenvalues are normalized to unity, representing PC1. The principal component with a normalized eigenvalue less than 0.1 does not contribute significantly.
It is evident that a larger number of principal components contribute to tidal deformability for most of the cases than those for the corresponding radius. For instance, consider a neutron star with a mass of $1.2M_\odot$; there are four significant principal components with normalized eigenvalues higher than 0.1 for tidal deformability, while only three for the radius.

In Fig. \ref{fig4}, we display the percentage of variation in neutron star properties explained by the first four principal components. The upper and lower panels illustrate the outcomes for uncorrelated (D1) and joint posterior (D2) distributions of nuclear matter parameters. The bars in green correspond to properties at a neutron star mass of $1.2M_\odot$, while the red bars represent $1.8M_\odot$. This figure depicts how much each PC contributes to capturing the variance. From PC1 to PC4, the portion of variance decreases. The total number of PCs needed to account for over 90\% of the variance is smaller for the radius than for the tidal deformability. As an example, consider $\Lambda_{1.2}$ in cases D1 and D2: to achieve over 94\% and 97\% variance, respectively, at least four PCs are necessary. On the other hand, for $R_{1.2}$, only three PCs are sufficient to achieve more than 94\% and 98\% variance in the respective cases.
\begin{figure*}[htp]
\centering
\includegraphics[width=0.9\textwidth,height=10cm]{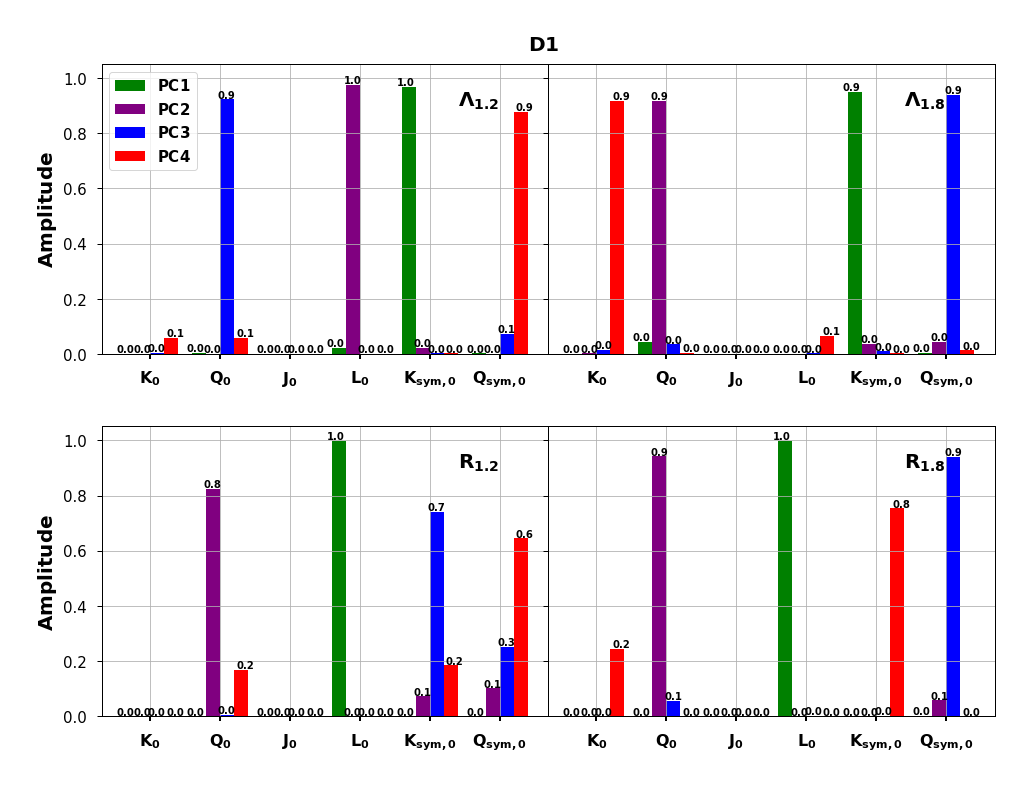}
\caption{(Color online) The square of the amplitude values of various nuclear matter parameters for each principal component. The results are presented for the uncorrelated nuclear matter parameter distributions (D1). The different PCs are indicated by different color bars.}\label{fig5}
\end{figure*}

\begin{figure*}[htp]
\centering
\includegraphics[width=0.9\textwidth,height=10cm]{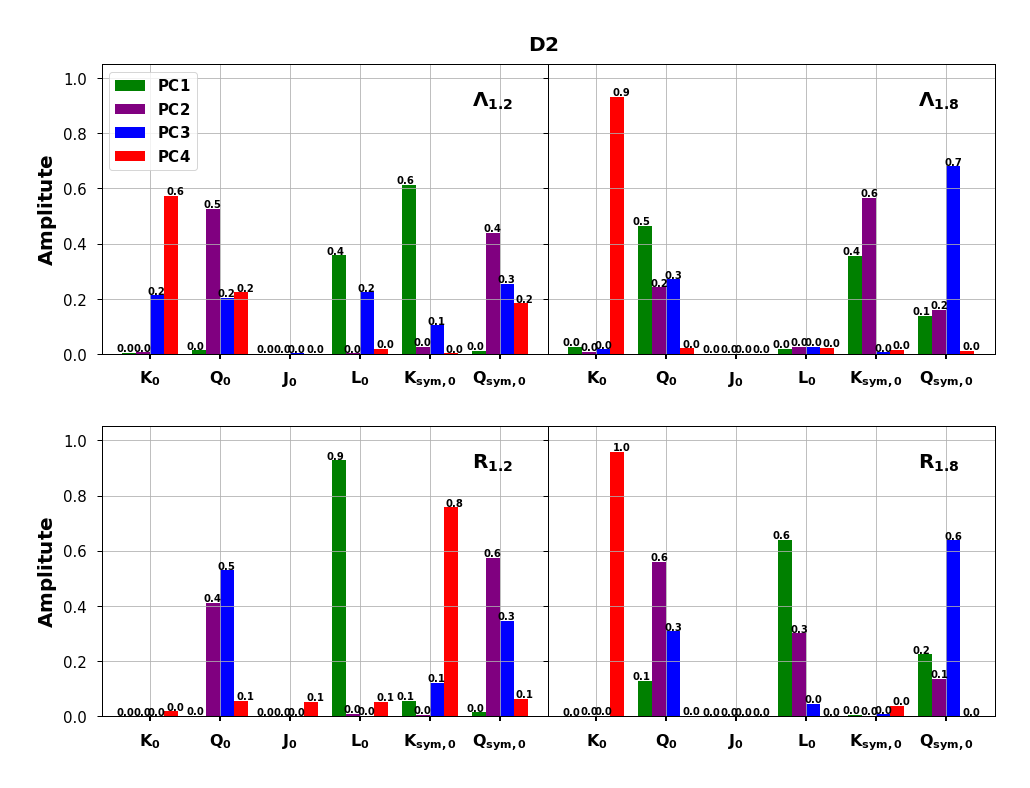}
\caption{(Color online) Same as Fig.\ref{fig5} but for  correlated nuclear matter parameters distributions (D2).}\label{fig6}
\end{figure*} 

\begin{figure*}[htp]
\centering   
\includegraphics[width=0.94\textwidth,height=14cm]{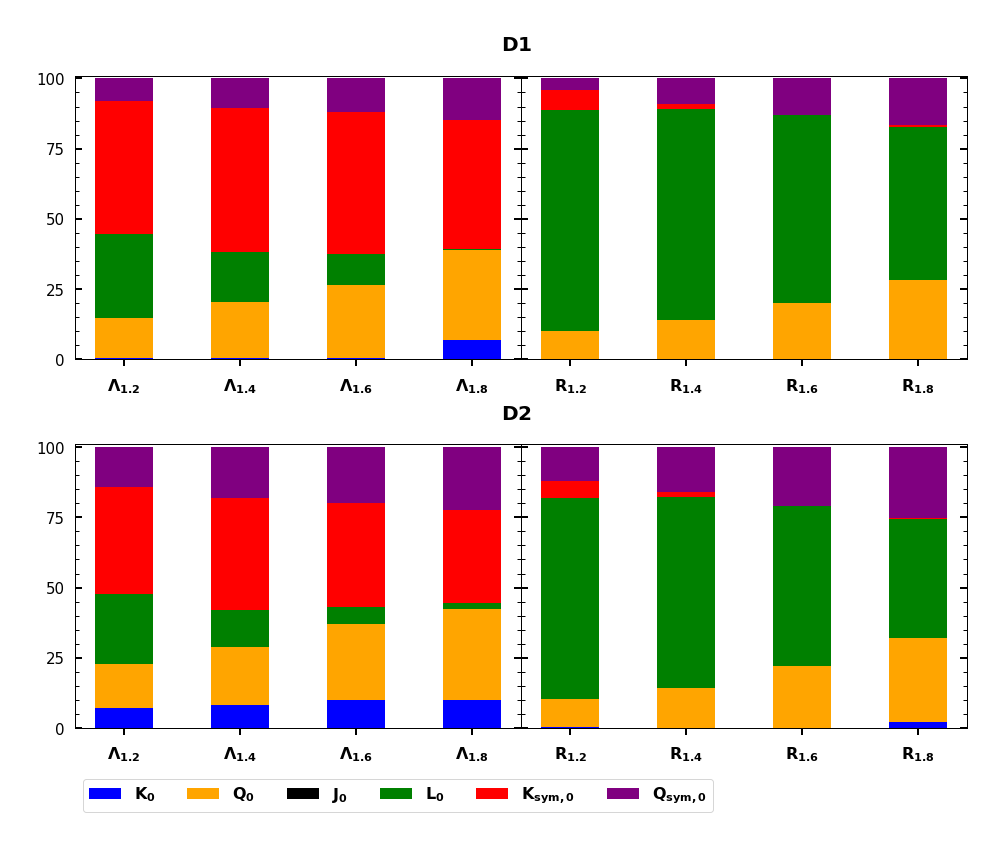}
\caption{(Color online) The values of the percentage contributions of nuclear matter parameters to the NS properties with masses range $1.2-1.8M_\odot$. The results in the upper panels and lower panels correspond to the D1 and D2 distributions of nuclear matter parameters, respectively.  The colors of the bar are related to the nuclear matter parameters.}\label{fig7}
\end{figure*}

The squared amplitude components of eigenvectors for a specific principal component provide insights into the contributions from different nuclear matter parameters. This is illustrated in Fig. (\ref{fig5}) and (\ref{fig6}), corresponding to D1 and D2 distributions, respectively. The upper panels pertain to tidal deformability, while the lower panels relate to radii. Various nuclear matter parameters contribute differently to the principal components, and this sensitivity depends on the chosen distribution of nuclear matter parameters. In the case of D1 distribution (Fig. \ref{fig5}), PC1 is primarily composed of a single parameter, K$_{\rm sym,0}$, for both $\Lambda_{1.2}$ and $\Lambda_{1.8}$. The PC2 is dominated by L$_0$ for $\Lambda_{1.2}$ and by Q$_0$ for $\Lambda_{1.8}$. 
The contribution to PC3 for $\Lambda_{1.2}$ is mainly  from Q$_0$, while for $\Lambda_{1.8}$, it is primarily Q$_{\rm sym,0}$. In PC4, Q$_{\rm sym,0}$ has the most significant contribution for $\Lambda_{1.2}$, while for $\Lambda_{1.8}$, it is  from K$_{0}$. Combining the outcomes of  Figures \ref{fig4} - \ref{fig6}, we can infer that the behavior of $\Lambda_{1.2}$ can be primarily explained by the linear combination of $K_{\rm sym,0}$, $L_0$, and $Q_0$ in the D1 distribution. However, in the case of the D2 distribution, the influence of $K_0$ becomes evident as an additional factor. Shifting the focus to $\Lambda_{1.8}$, the impact of $L_0$ diminishes, while the contributions from iso-scalar parameters like $K_0$ and $Q_0$, along with the iso-vector parameter $Q_{\rm sym,0}$, increase. This suggests that $\Lambda_{1.8}$ is
influenced by a simultaneous interplay of various nuclear matter parameters.  For radii (R$_{1.2}$ and R$_{1.8}$), PC1 is dominated by L$_0$, while PC2 by Q$_0$. In PC3 for R$_{1.2}$, contributions from K$_{\rm sym,0}$ and Q$_{\rm sym,0}$ are notable, whereas for R$_{1.8}$, it's mainly Q$_{\rm sym,0}$.
In Fig. \ref{fig6}, similar outcomes are depicted, but for D2 distributions, which are notably different from those in Fig. \ref{fig5}. Multiple nuclear matter parameters contribute to most PCs due to correlations in the D2 distribution. Strong correlations, such as between L$_0$ and K$_{\rm sym,0}$, lead to their combined contributions. For instance, in PC1, both $\Lambda_{1.2}$ and R$_{1.2}$ are influenced by L$_0$ and K$_{\rm sym,0}$, while $\Lambda_{1.8}$ is influenced by Q$_0$, K$_{\rm sym,0}$, Q$_{\rm sym,0}$  and R$_{1.8}$ by Q$_0$, L$_{0}$, and Q$_{\rm sym,0}$. The PC2's contributions include Q$_0$ and Q$_{\rm sym,0}$ for $\Lambda_{1.2}$ and R$_{1.2}$, Q$_0$, K$_{\rm sym,0}$, and Q$_{\rm sym,0}$ for $\Lambda_{1.8}$, and Q$_0$, L$_0$, and Q$_{\rm sym,0}$ for R$_{1.8}$. The PC3 encompasses all NMPs except J$_0$ for $\Lambda_{1.2}$, Q$_0$ and Q$_{\rm sym,0}$ for $\Lambda_{1.8}$, Q$_0$, K$_{\rm sym,0}$, and Q$_{\rm sym,0}$ for R$_{1.2}$, and Q$_0$, Q$_{\rm sym,0}$ for R$_{1.8}$. Similarly, PC4 involves K$_0$, Q$_{0}$, and Q$_{\rm sym,0}$ for $\Lambda_{1.2}$, while K$_0$ dominates PC4 for $\Lambda_{1.8}$. The contributions of PC4 for radii are negligible.

We computed the total contributions for each of the nuclear matter parameters by summing their contributions to each of the PCs.  The weights for this calculation are derived from the product of the square of amplitude for each nuclear matter parameter and the corresponding eigenvalues of the PCs listed in Table \ref{tab2}. These weights are then normalized so that the total sum of contributions from all nuclear matter parameters equals unity. The percentages representing the contributions of individual nuclear matter parameters to neutron star properties within the mass range of $1.2-1.8M_\odot$ are shown in Fig. \ref{fig7}.  The upper and lower panels show the results for D1 and D2 distributions, respectively. Different colors are used to depict the percentage contributions of each nuclear matter parameter. It is important to note that the contributions of different nuclear matter parameters to neutron star properties are influenced by the choice of nuclear matter parameter distributions. For both D1 and D2 distributions, the contributions of specific nuclear matter parameters to tidal deformability and radius can differ by up to 20\%. However, when categorized broadly into iso-scalar and iso-vector parameters, these differences diminish. Iso-scalar parameters,   $K_0$ and $Q_0$, contribute together, while the remaining parameters contribute to the iso-vector category. The total contributions from iso-scalar parameters increase, while it decreases accordingly for the iso-vector parameters with the increase in NS mass.
For instance, the  contribution from iso-scalar parameters in the case
of D1 (D2) distributions increases  from 15\% to 40\% (24\% to  44\%)
for tidal deformability with the increase in NS mass from 1.2$M_\odot$ to
1.8$M_\odot$.  Concerning radius, iso-scalar parameter contributions
increase from approximately 10\% to 30\% as the neutron star mass
increases from $1.2M_\odot$ to $1.8M_\odot$.

{Finally, it may be emphasized that the present work highlights the necessity of multivariate analysis of neutron star properties through PCA as one of the tools. A more comprehensive investigation addressing improved treatment of crust EoSs and high-density EoSs is warranted for a more realistic assessment \cite{Carreau:2019zdy}.}

\section{Summary and Outlook} \label{summary}
We have addressed an unresolved issue of connecting the nuclear matter parameters to the key neutron star properties, such as tidal deformability and radius. 
The outcomes of the majority of the investigations exploring the correlations between properties of neutron stars and individual nuclear matter parameters, which describe the equations of state, are at variance.  We have exploited the efficacy of Principal Component Analysis, a sophisticated analytical tool, in order to establish a comprehensive connection between multiple nuclear matter parameters and the key properties of neutron stars, with the ultimate aim of shedding some light on the existing issue. 
The EoSs essential for describing neutron star matter within the core region up to a density of $2\rho_0$ have been derived. This was accomplished by utilizing both uncorrelated uniform and joint posterior distributions of nuclear matter parameters. To ensure continuity and consistency, each of these distinct EoSs is joined smoothly at $2\rho_0$ by a diverse set of the EoSs obtained by parameterizing the speed of sound such that it remains causal and approaches the conformal limit gradually.

We have found that the variability in the considered neutron star properties requires the incorporation of more than one principal component. These observations emphasize that the neutron star properties depend on multiple nuclear matter parameters. In particular, tidal deformability demands the inclusion of three or more principal components to account for over 90\% of its variations, while for the radius, two principal components suffice to explain similar variations.   
The dominance of nuclear matter parameters contributing to the principal components depends on the specific NS properties and their mass.
For instance, in the case of the tidal deformability of a neutron star with a mass of 1.2$M_\odot$, the symmetry energy curvature parameter $K_{\rm sym,0}$ emerges as the primary contributor to the first principal component.  The second and third principal components are significantly influenced by the symmetry energy slope parameter $L_0$ and the skewness parameter $Q_0$ for symmetric nuclear matter, respectively. The significance of iso-scalar nuclear matter parameters, specifically the incompressibility coefficient K$_0$ and the skewness parameter $Q_0$ of symmetric nuclear matter, becomes more pronounced with an increase in the mass of the neutron star. When the NS mass reaches 1.8$M\odot$, the incompressibility coefficient $K_0$ surpasses the importance of the symmetry energy slope parameter $L_0$. When considering the radius of a neutron star at lower masses, the symmetry energy slope parameter $L_0$ stands out as the primary driver behind the observed variations. Overall, when analyzing the collective impact of iso-scalar parameters ($K_0$ and $Q_0$), these contributions exhibit an approximately 25\% increase with the increase in neutron star mass from 1.2$M_\odot$ to 1.8$M_\odot$.

The properties of neutron stars are the composite functions of the
nuclear matter parameters, primarily governing the equation of state
within the density range of approximately 0.5 to 2 times the nuclear
saturation density. Recent studies \cite{Imam:2023ngm, Richter:2023zec} indicate that the tidal deformability and radii of neutron stars are sensitive to not only individual nuclear
matter parameters but also to fractional and higher integral powers of
these parameters, as well as their products.  Consequently, there exists an intricate correlation between neutron star properties and nuclear matter parameters, characterized by their complexity.  These complex correlations can be further explored by extending the current investigation to incorporate non-linear contributions derived from individual parameters or from their combinations.                                                                                                                                                        
\section{Acknowledgements} 
The authors would like to thank Tuhin Malik and Sk Md  Adil Imam for a careful reading of the paper and important suggestions. NKP would like to acknowledge the Department of Science and Technology, Ministry of Science and Technology, India, for the support of DST/INSPIRE Fellowship/2019/IF190058. BKA acknowledges partial support from the Department of Science and Technology, Government of India with grant no.
CRG/2021/000101.

\bibliographystyle{apsrev4-1}
\bibliography{main}

\setcounter{equation}{0}
\setcounter{figure}{0}
\setcounter{table}{0}
\setcounter{page}{1}
\makeatletter
\renewcommand{\theequation}{S\arabic{equation}}
\renewcommand{\thefigure}{S\arabic{figure}}
\renewcommand{\thetable}{S\arabic{table}}

\widetext
\begin{minipage}{\textwidth}
\centering .
\appendix
\section{Computational details of PCA Analysis}\label{method-pca}
\end{minipage}

The technique of  Principal Component Analysis (PCA) is commonly used in data science for feature extraction and dimensionality reduction for a given target variable. Here are the steps to implement the PCA to identify key features
associated with a given target variable.

\begin{enumerate}
    \item Data Preprocessing: The initial step involves the preparation of the dataset.  The dataset should contain both the target variable and the features. 
    \item Standardization: The data should be standardized by adjusting it to have a mean of zero and a variance of unity. This stage is crucial for ensuring that all features are standardized to a similar scale, preventing any particular feature from exerting undue influence on the PCA process due to its greater magnitudes.
    
    \item The Covariance Matrix: Calculate the covariance matrix for the standardized dataset. The covariance matrix is a mathematical representation that captures both the variances and connections among different features. The computational complexity of the covariance matrix is $O(ND \times min(N, D))$, which results by multiplying two matrices of size $D \times N$ and $N \times D$, respectively. Here, $N$ is the number of samples, and $D$ is the dimensionality or simply the number of features.
    
    \item Eigenvalue decomposition: Determine the eigenvectors and eigenvalues of the covariance matrix. The eigenvectors represent the PCs, while the eigenvalues indicate the amount of variance explained by each PC. Arrange the eigenvalues in descending order to assign higher priority to the principal components that account for the greatest amount of variance. 
    
    \item Selection of PCs: Determine the number of main components that will be kept. The selection of this option depends upon our target to reduce dimensionality.  We can either select a certain number of top PCs or opt to keep a given proportion of the total variance (e.g., 95\%).
    \item Amplitude: An analysis is conducted to evaluate the square of the amplitude values of the feature on each PC. Features that have larger absolute loadings on a specific PC are regarded as having a greater contribution to that component.
    \item Percentage Contribution: The final step is to find the percentage contributions of all features to the target variable. The total contributions of a given feature are obtained by summing their contributions from each of the PCs weighted by the corresponding normalized eigenvalue. 
\end{enumerate}
The overall complexity of the PCA analysis is $O(ND\times min(N, D))$.
By following the above steps, One can identify the importance of PCs by reducing the dimensionality of the data and can extract the key parameters from the dataset for a specific target.


\end{document}